\documentclass[12pt]{article}
\usepackage{amsmath}
\usepackage{amssymb}
\usepackage{fullpage}

\def\##1{{\bf #1}}
\def\=#1{\underline{\underline #1}}

\def\les{\left[}
\def\ris{\right]}

\def\ric{\right\}}

\def\c#1{\cite{#1}}

\def\r#1{(\ref{#1})}

\def\eps{\epsilon}
\def\epso{\epsilon_0}
\def\muo{\mu_0}
\def\ko{k_0}

\def\tde{\tan\,\delta_\epsilon}
\def\tdm{\tan\,\delta_\mu}

\begin{document}

\noindent{\large\bf REVERSED CIRCULAR DICHROISM OF \\ ISOTROPIC CHIRAL MEDIUMS WITH NEGATIVE REAL \\ PERMEABILITY AND PERMITTIVITY }

\vskip 1.0 cm

\noindent{\bf Akhlesh Lakhtakia}$^1$
\vskip 0.4 cm

\noindent $^1$CATMAS --- Computational \& Theoretical Materials Sciences Group\\
\noindent Department of Engineering Science and Mechanics\\
\noindent Pennsylvania State University\\
\noindent University Park, PA 16802--6812
\vskip 1 cm

\noindent {\bf ABSTRACT:}
Negative real parts of the permittivity and permeability lead an isotropic chiral medium to
exhibit
circular dichroism that is reversed with respect to that exhibited by an identical
medium but the real parts of whose permittivity and permeability are positive.


\section{Introduction}
The Drude--Born--Fedorov constitutive relations of a homogeneous,
isotropic chiral medium (ICM) are stated as
\c{Lak1}
\begin{equation}
\left. \begin{array}{cccc}
\#D(\#r,\omega) = \eps(\omega) \les \#E(\#r,\omega) +
\beta(\omega)\, \nabla\times\#E(\#r,\omega)\ris \\
\#B(\#r,\omega) = \mu(\omega) \les \#H(\#r,\omega) +
\beta(\omega)\, \nabla\times\#H(\#r,\omega)\ris
\end{array}\ric \, .
\end{equation}
The permittivity of the ICM
is denoted by $\eps(\omega)$, and its permeability by $\mu(\omega)$,
whilst $\beta(\omega)$ is the chirality parameter. 
An $\exp(-i\omega t)$ time--dependence is implicit
throughout and the dependences on the angular frequency
$\omega$ are understood from here onwards. 

When a linearly
polarized plane wave is normally incident on an ICM slab
of infinite lateral dimensions, the transmitted plane wave is elliptically
polarized with its vibration ellipse rotated with respect to direction of the incident
electric field phasor \c{Pye, LakSingh}. These two effects are quantified via
the optical rotation $\delta$ and the circular dichroism $\psi$, expressed on
a per--unit--thickness basis as \c{Char}
\begin{equation}
\left. \begin{array}{ccc}
\delta = {\rm Re}\les \gamma_1 - \gamma_2\ris/2\\
\psi = {\rm Im}\les \gamma_1 - \gamma_2\ris/2
\end{array}\ric\,,
\end{equation}
where the wavenumbers
\begin{equation}
\label{gam}
\left. \begin{array}{cccc}
\gamma_{1} = \frac{\omega\sqrt{(\eps\mu)}}{1- \omega\sqrt{(\eps\mu)}\beta}\\[7pt]
\gamma_{2} =  \frac{\omega\sqrt{(\eps\mu)}}{1+\omega\sqrt{(\eps\mu)}\beta}
\end{array}\ric \, ;
\end{equation}
hence,
\begin{equation}
\label{dp}
\delta +i\psi =\frac{\omega^2\eps\mu}{1-\omega^2\eps\mu\beta^2}\,\beta\,.
\end{equation}
A related quantity of interest is the impedance
\begin{equation}
\eta = \sqrt{(\mu/\eps)}\,,
\end{equation}
which appears in the decomposition of the electric and the magnetic
field phasors in terms of Beltrami fields \c{Lak1}.

Typically, $\beta$ is a small parameter that appears to manifest chiefly
via $\delta$ and $\psi$.
The influence of $\beta$ is immediately obvious
from \r{gam} and \r{dp}: a change in its sign results in the interchange of
the values of the two wavenumbers, which thus reverses the signs of both the optical rotation
and the circular dichroism \c{Lak1}. 
Pairs of mediums differing only in $\beta$ and/or in $\eta$
have been contrasted earlier in this journal \c{Lak2,Lak3}.

The objective of the present communication is to present the
observable consequences of a very recent development
on isotropic chiral mediums:  Shelby {\em et al.\/} have
constructed and tested a dielectric/magnetic composite material which
appears to
possess permeability and permittivity with negative real parts in a
particular frequency range \c{SSS}. Although their material
is not exactly as advertised, future possibilities mandate
the examination of observable repercussions on ICMs. 

\section{Analysis}
Let us consider electromagnetic field phasors in an ICM at a frequency
that does not lie in an absorption band 
\c{Char}. Accordingly,
\begin{equation}
\label{defem}
\left.\begin{array}{cc}
\eps = a\, (\pm 1 +i \,\tde)\, \epso \\
\mu = b \,(\pm 1 + i\,\tdm) \,\muo
\end{array}\ric\,,
\end{equation}
where $a > 0$, $b>0$, $\tde\geq 0$ and $\tdm\geq 0$ are real--valued,
and we impose the condition of low loss (i.e., ${\rm Im}\les \gamma_{1,2}\ris \leq
{\rm Re}\les \gamma_{1,2}\ris$). The upper signs in the foregoing equations
hold for the {\em normal\/} case, the lower for the case suggested by the
experimental results of Shelby {\em et al.\/} \c{SSS}. The permeability
and the permittivity of free space are denoted, respectively, by $\muo$ and $\epso$.

Correct to the first order in both $\tde$ and $\tdm$, we obtain
\begin{equation}
\gamma_1\approx \frac{\ko m}{1-\ko m \beta}\,
\les 1 \pm \frac{i}{1-\ko m \beta}\,\frac{\tdm+\tde}{2}\ris\,
\end{equation}
and
\begin{equation}
\gamma_2\approx\frac{\ko m}{1+\ko m \beta}\,
\les 1 \pm \frac{i}{1+\ko m \beta}\,\frac{\tdm+\tde}{2}\ris\,
\end{equation}
from \r{gam} and \r{defem},
where
$\ko = \omega\sqrt{(\epso\muo)}$ is the free--space wavenumber
and $m = +\sqrt{(ab)}$. Therefore, the right side of \r{dp}
can be approximated as follows:
\begin{equation}
\label{dp1}
\delta+i\psi
\approx \frac{\ko^2m^2}{1-\ko^2m^2\beta^2}\,
\les 1 \pm i\, \frac{\tdm+\tde} {1-\ko^2m^2\beta^2}\ris\,\beta\,.
\end{equation}
In the same way,
\begin{equation}
\label{eta1}
\eta \approx  \sqrt{(b/a)}\,
\les 1 \pm i \,\frac{\tdm-\tde}{2}\ris\,\eta_0\,,
\end{equation}
where $\eta_0 = \sqrt{(\muo/\epso)}$ is the intrinsic
impedance of free space.

\section{Conclusions}
Equation \r{dp1} shows that the optical rotation remains
unaffected if the real parts of both the permittivity and the
permeability change signs, but the circular dichroism reverses
in sign. This conclusion holds provided ${\rm Im}\les
\beta\ris$ can be ignored, i.e, at frequencies away from a circular dichroism band
\c{Char}.
Otherwise, optical rotation is also affected. Equation \r{eta1} shows that 
only the imaginary part of $\eta$
changes sign, if the signs of  the real parts of both the permittivity and the
permeability are altered.

Suppose two ICMs labeled {\em p} and
{\em q} possess identical response properties at a
certain frequency~---~except that $0<{\rm Re}\les\eps_p\ris = - {\rm Re}\les\eps_q\ris$ and
$0<{\rm Re}\les\mu_p\ris = -{\rm Re}\les\mu_q\ris$. Correct to the zeroth
order in both $\tde$ and $\tdm$, the two ICMs are
isoimpedant as well as isorefractive; yet, they could be distinguished
from each other by employing obliquely incident plane waves.
Of course, \r{dp1} and \r{eta1} show that the two ICMs are neither
isoimpedant nor isorefractive, correct to the first order in $\tde$ and $\tdm$.

\end{document}